\documentclass[a4paper,10pt,prl,twocolumn,superscriptaddress,nofootinbib]{revtex4-1}

\usepackage[dvips]{graphicx}
\usepackage[dvips,usenames]{color}
\usepackage{amsmath}
\usepackage{amsfonts}
\usepackage{txfonts}
\usepackage{amssymb}
\usepackage{amstext}
\usepackage[sort&compress]{natbib}
\usepackage{graphics,graphicx}

\newcommand{\be}{\begin{equation}}
\newcommand{\ee}{\end{equation}}
\newcommand{\bea}{\begin{eqnarray}}
\newcommand{\eea}{\end{eqnarray}}

\newcommand{\curly}[1]{\left\{{#1}\right\}}
\newcommand{\crc}[1]{\left({#1}\right)}

\begin{document}

\title{Surprising Interactions of Markovian noise and Coherent Driving}

\author{Shai Machnes}
\email{shai.machnes@weizmann.ac.il}
\affiliation{Weizmann Institute of Science, Israel}
\author{Martin Plenio}
\affiliation{Institut f{\"u}r Theoretische Physik, Universit{\"a}t
Ulm, D-89069 Ulm, Germany}

\begin{abstract}
We derive the explicit commutation relations for the generators of
quantum dynamical semigroup - Markovian superoperator evolution, allowing
the extension of Baker-Campbell-Hausdorff-type relations to general
Lindblad-type evolutions. This provides a novel tool for exploring
the interaction of time-dependent coherent and Markovian evolutions - a surprisingly rich set of behaviors which include
deformation by coherent driving of Markovian terms, generation of new Lindblad terms from time-dependent noise and even a coherent
driving term generated by the interaction of time-separated Markovian noises. Finally, we consider the Lindblad superoperators as vectors in a higher space, allowing us to extract the infinitely divisible subspace of a given channel and to recover its Lindblad form.
\end{abstract}

\maketitle

\subsection*{Introduction}

The importance of quantum effects in future technology cannot be overstated,
whether desired (quantum computation, metrology, etc) or detrimental
(leakage current in MOSFETs).

For some applications, one desires to isolate the quantum system and
protect it from dephasing or relaxation resulting from uncontrolled
interaction with the environment. To that end, a toolbox of methodologies
have been developed to allow the stabilization of quantum states in
the presence of decoherence and to implement coherent operations in
a noisy environment. The tools range from the design of decoherence-free
subspacs \cite{Lidar review of DFS and DD,DFS,DFS 2} coupled with
a measurement-driven Zeno-type controls to keep the state within the
protected subspace \cite{Zeno on DFS}, to more dynamic methods such
as pulsed/dynamical-decoupling \cite{Dynamical decoupling,pulsed-decoupling,BB2,Lidar review of DFS and DD} and optimal control \cite{OC1,OC2,OC3,OC4,OC5,OC6,OC7,OC8,Optimal control Markovian}. On the other-hand, non-coherent effects resulting from environment
interaction may be positively utilized in a variety of tasks, such
as state preparation \cite{State prep with noise 1,State prep with noise 2,State prep with noise 3},
quantum map synthesis \cite{map prep}, enhancement energy transfer
in photosynthetic systems \cite{photosynth1,photosynth2,photosynth3,photosynth4}. 
In all these cases, there is strong interplay between coherent and
incoherent dynamics.

Surprisingly, even in the most commonly used model for environment interactions, that of Markovian interactions,
the algebra governing the compound effect of consecutive periods of coherent driving and Markovian noise has yet to be fully worked-out. It is the goal of this paper to provide algebraic tools, with clear physical interpretation, to allow deep examination of the interplay of coherent and incoherent state evolution.

This letter is organized as follows: First we review the Lindblad-Kossakowski
master equation, describing quantum Markovian semigroup evolution,
and the Baker-Campbell-Haussdorf (BCH) relations, which allow us to
examine the interaction of driving pulses amongst themselves and with
the free evolution of the system. We shall introduce a new notation
which will allow us to derive explicit commutations relations of the
dynamical semigroup generators, thus promoting the BCH relations to
superoperator maps. We shall then use this to examine the intricacies
and richness of interactions between Markovian noises and coherent
driving, culminating in an example where two Markovian
noise actions, separated by time, interact to generate a wholly coherent
term. Finally, we view superoperators as vectors in a higher ("super-super") operator
Hilbert space, and utilize the language of bi-orthogonal bases to present a method
of projecting a general map onto the sub-space of infinitely divisible Markovian superoperators
and further phrasing the projected map in its Lindblad form.
Some concluding remarks close-out the discussion.

\subsection*{Background Review}
\subsubsection*{The Lindblad-Kossakowski quantum master equation}
Coherent evolution in quantum mechanics, as described by the Schr{\"o}dinger
equation $i\hbar\partial_{t}\psi=H\psi$, defines a set of continuous,
one-parameter, exponential unitary maps $\mathcal{U}\left(t\right)=\exp\left(-\frac{i}{\hbar}tH\right)$,
identifying the elements of the Lie algebra with the infinitesimal
generator of the group, the Hamiltonian.

A general Markovian (memory-less) time-homogeneous, trace-preserving
and completely positive evolution of any open quantum system is described
by the Lindblad-Kossakowski master equation (henceforth LKME)
\cite{Textbook - Breuer}
\begin{equation}\displaystyle
\dot{\rho}=\mathcal{H}\left(H\right)+\mathcal{L}\left(\Gamma,L_{1}\ldots L_{M}\right),
\end{equation}
\begin{equation}\displaystyle
\mathcal{H}\left(H\right)=-\frac{i}{\hbar}\left[H,\rho\right],
\end{equation}
\begin{equation}\displaystyle
\mathcal{L}\left(\Gamma,L_{1}\ldots L_{M}\right)=\sum_{j,k}\Gamma_{k,j}\left(L_{k}\rho L_{j}^{\dagger}-\frac{1}{2}\left(\rho L_{j}^{\dagger}L_{k}+L_{j}^{\dagger}L_{k}\rho\right)\right)
\label{eq:NonCanonicalLinbdblad}
\end{equation}
where the $\Gamma$ matrix is Hermitian and positive semidefinite.

Master equations of this form are in very common use in a great number
of fields, including quantum optics, semiconductor physics, NMR, decay
of unstable systems, thermalization, Brownian motion, etc. Such an
equation describes the irreversible evolution of quantum system, weakly
interacting with a stationary environment in such a way that the timescale
of system dynamics and relaxation is much longer than decay time of
correlations with and in the environment, so that one may ignore memory
(non-Markovian) effects.

$\Gamma$, above, may can be diagonalized, $\gamma=U_{\Gamma}\Gamma U_{\Gamma}^{\dagger}$,
with $\gamma$ being a diagonal positive semidefinite matrix. One
then defines $A_{j}=\sum_{j,k=1}^{N^{2}-1}\left(U_{\Gamma}^{\dagger}\right)_{k,j}L_{k}$
, resulting in
\begin{equation}\displaystyle
\dot{\rho}=\mathcal{H}\left(H\right)+\sum_{j}\gamma_{j}\mathcal{L}_{1}\left(A_{j}\right),
\end{equation}
\begin{equation}\displaystyle
\mathcal{L}_{1}\left(A\right)=A\rho A^{\dagger}-\frac{1}{2}\left(\rho A^{\dagger}A+A^{\dagger}A\rho\right).
\end{equation}
One may directly show that $\mathcal{L}_{1}\left(A\right)=\mathcal{L}_{1}\left(A-\left(trA\right)\mathcal{I}\right)+\mathcal{H}\left(\frac{i\hbar}{2}\left(\left(trA^{\dagger}\right)A-\left(trA\right)A^{\dagger}\right)\right)$,
and therefore we remove the trace of the Lindblandian operators $A$
into the Hamiltonian, leaving all such operators trace-less. Trivially, one may also remove the trace from the Hamiltonian, $H\longrightarrow H-\left(trH\right)\mathcal{I}$ without altering system dynamics. This is considered the LKME canonical form \cite{Hall Canonical}.

We may consider $\mathcal{H}$, $\mathcal{L}$ and $\mathcal{L}_{1}$
to be superoperators acting on the space of density matrices; and
the set of continuous, completely positive, one-parameter maps defined
by the LKME form a one-parameter dynamical Lie semigroup \cite{Lie semigroups Dirr},
with$\mathcal{H}$, $\mathcal{L}$ and $\mathcal{L}_{1}$ serving
as the infinitesimal superoperator generators.

\subsubsection*{The Lamb shift}
In the derivation of the LKME, one performs several approximations and makes several assumptions: the absence of correlations between system and environment in the initial state, weak coupling (leading to the state remaining a product state; Born), memory-less nature of the evolution (dependency only on the current state of the system and not on past states; Markov) and in some cases the secular approximation (RWA (Rotating Wave Approximation) - averaging over rapidly oscillating terms). In addition, a coherent term, resulting from the system-bath interaction, is added to the Hamiltonian - the Lamb shift term \cite{Textbook - Breuer}.

Explicitly, given a general interaction Hamiltonian,
\begin{equation}\displaystyle
H_I = \sum_k A_k \otimes B_k = \sum_{k} \crc{\sum_\omega A_k \crc{\omega}} B_k
\end{equation}
with $A_k,\ B_k$ being, respectively, system and bath Hermitian operators, and $A_k =\sum_\omega A_k \crc{\omega}$ being the energy-basis decomposition of the system operators.  After performing the secular approximation we end up with the LKME in non-canonical form
\begin{equation}\displaystyle
\displaystyle
\begin{array}{c}
\dot{\rho}=\mathcal{H}\left(H_{0}+H_{L.S.}\right)+\sum_\omega \mathcal{L}\crc{\Gamma, A_1 \crc{\omega}, A_2 \crc{\omega}, \ldots } \\
H_{L.S.}=\sum_{j,k,\omega}S_{j,k}\crc{\omega}A_{j}^{\dagger}\crc{\omega}A_{k}\crc{\omega} \\
\Gamma=\frac{1}{2} \crc{R+R^\dagger} \\
S=\frac{1}{2} \crc{R-R^\dagger}
\end{array}
\label{eq:HLS}
\end{equation}
with $R$ being the one-sided Fourier transform of the bath auto-correlation function, $R_{j,k}\crc{\omega}=\int_{0}^{\infty}d\tau e^{i\omega\tau}\left\langle B_{j}^{\dagger}\left(\tau\right)B_{k}\left(0\right)\right\rangle$. Note that the Lamb shift Hamiltonian, $H_{L.S.}$, has been explicitly removed from the Lindbladian terms. As we shall see below, it makes a surprise reappearance in the interaction of time-separated noise terms.

\subsubsection*{The Baker-Campbell-Haussdorf and Wei-Norman equations}
In the context of quantum optimal control, one implements the desired
unitary linear map by concatenating multiple pulses, intertwined with
periods of free evolution. To ascertain analytically the total effect
of such a pulse-train, one utilizes the Baker-Campbell-Haussdorf (BCH)
relation \cite{BCH-1,BCH-2,BCH-3}, which are essentially a matrix identity

\begin{multline}
\exp\left(X\right)\exp\left(Y\right)=\,\,\,\,\,\,\,\,\,\,\,\,\,\,\,\,\,\,\,\,\,\,\,\,\,\,\,\,\,\,\,\,\,\,\,\,\,\,\,\,\,\,\,\,\, \\
\exp\left(X+Y+\frac{1}{2}[X,Y]+\frac{1}{12}[X,[X,Y]]-\frac{1}{12}[Y,[X,Y]]+\dots\right).
\label{eq:BCH}
\end{multline}

Some conditions are required to ensure the existence of a convergence radius of the exponential
map, and are discussed in \cite{BCH convergence radius}.
When utilizing the BCH relations, we shall restrict ourselves to the finite-dimensional case.
One may extend the relations to any number of exponents \cite{BCH generalization},
where the series is represented as a summation over products of the
exponent generators (and while by induction it is clear said series
can be expressed in terms of commutators, as in eq. (\ref{eq:BCH})
above, to our knowledge the general form has not yet been explicitly
reformulated). The Wei-Norman relations \cite{Wei Norman} stipulate
the opposite - that an exponential map of a sum may be represented
as a product of exponential maps of algebra's generators.

From a physical perspective, given a series of unitary evolutions
each generated by a time-independent Hamiltonian, the BCH relations
provide us with a way of phrasing the overall evolution with a single
time-independent Hamiltonian. And while often the trajectory of the
state in the LHS and RHS of eq. $\left(\ref{eq:BCH}\right)$ will
coincide only at the initial and final times, the explanatory power
of such a re-phrasing is significant (see, for example \cite{superfast,superfastCavity}
where impulsive driving is intertwined with free evolution, to generate
novel interactions by the commutator of the pulse and free Hamiltonian).

Stressing the memory-less nature of the classical Markov process, and following
the definitions in \cite{Rivas Huelga,Plenio review,Div. Q. Chann - Wolf and Cirac},
one can take the position that a sequence of quantum Markovian evolutions is itself a Markovian evolution.
Our goal then is to lift the BCH relations from unitary to semigroup evolution, from operator to superoperator.
With this we will be able to understand the unique nature of a time-dependent Markovian process, in that, unlike time-dependent coherent evolution, the compound evolution takes a somewhat different form than the time-independent one.

\subsection*{Matrix superoperators and vec-ing of the density matrix}

To explicitly formulate the commutation relations of the LKME generators,
we shall introduce the following notation for superoperators: given
a left and right operators acting on a density matrix $L\rho R$,
one may transform the matrix $\rho$ into a column vector, row-first,

\begin{equation}\displaystyle
\vec{\rho}:=\left(\rho_{1,1},\rho_{1,2}\ldots\rho_{1,N},\rho_{2,1},\rho_{2,2}.\ldots,\rho_{N,N}\right)^{T}.
\end{equation}

This procedure is known as vec-ing \cite{Textbook - Horn & Johnson,alt veccing}. Next, we shall define $\odot$ in
the following manner
\begin{equation}\displaystyle
\overrightarrow{L\rho R}=\left(L\otimes\left(R^{T}\right)\right)\vec{\rho}=:\left(L\odot R\right)\vec{\rho}.
\end{equation}

We define the superoperator equivalents of $\mathcal{H}$ and $\mathcal{L}$,

\begin{equation}\displaystyle
\mathcal{H}^{s}\left(H\right):=\frac{-i}{\hbar}\left(H\odot I-I\odot H\right)
\end{equation}
and
\begin{equation}\displaystyle
\mathcal{L}_{1}^{s}\left(A\right):=A\odot A^{\dagger}-\frac{1}{2}I\odot A^{\dagger}A-\frac{1}{2}A^{\dagger}A\odot I.
\end{equation}

Allowing us to rewrite the LKME in explicit superoperator notation
\begin{equation}\displaystyle
\dot{\vec{\rho}}=(\mathcal{H}^{s}\left(H\right)+\sum_{k}\gamma_{k}\mathcal{L}_{1}^{s}\left(A_{k}\right))\vec{\rho}
\end{equation}

defining the semigroup

\begin{equation}\displaystyle
\vec{\rho}\left(t\right)=e^{t(\mathcal{H}^{s}\left(H\right)+\sum_{k}{\gamma_{k}\mathcal{L}^{s}_{1}\left(A_{k}\right)})}\vec{\rho}.
\end{equation}

\subsection*{Explicit form of semigroup generator algebra }

Noting that $\left(A\odot B\right)\left(C\odot D\right)=AC\odot DB$
we can now arrive at the commutation relations for the LKME generators
(derivation is straightforward, if somewhat cumbersome):

\begin{eqnarray*}
\left[\mathcal{H}^{s}\left(H\right),\mathcal{H}^{s}\left(G\right)\right] & = & \mathcal{H}^{s}\left(-\frac{i}{\hbar}\left[H,G\right]\right),
\end{eqnarray*}

\begin{multline}
\left[\mathcal{H}^{s}\left(H\right),\mathcal{L}^{s}\left(A\right)\right]=\mathcal{L}^{s}\left(\Gamma=-\frac{i}{\hbar}\left(\begin{array}{cc}
0 & -1\\
1 & 0
\end{array}\right),\left\{ A,\left[H,A\right]\right\} \right)\\
=-\frac{1}{2\hbar}\mathcal{L}_{1}^{s}\left(A+i\left[H,A\right]\right)+\frac{1}{2\hbar}\mathcal{L}_{1}^{s}\left(A-i\left[H,A\right]\right)\label{eq:HL-comm}
\end{multline}

and

\begin{multline}
\left[\mathcal{L}^{s}\left(A\right),\mathcal{L}^{s}\left(B\right)\right]=\,\,\,\,\,\,\,\,\,\,\,\,\,\,\,\,\,\,\,\,\,\,\,\,\,\,\,\,\,\,\,\,\,\,\,\,\,\,\,\,\,\,\,\,\,\,\,\,\,\,\,\,\,\,\,\,\,\,\,\\
=\mathcal{L}_{1}^{s}(AB)-\mathcal{L}_{1}^{s}(BA)+\,\,\,\,\,\,\,\,\,\,\,\,\,\,\,\,\,\,\,\,\,\,\,\,\,\,\,\,\,\,\,\,\,\,\,\,\,\,\,\,\\
+\frac{1}{2}\mathcal{L}^{s}\left(\Gamma=\left(\begin{array}{cc}
0 & 1\\
1 & 0
\end{array}\right),\{B,\left[B,A^{\dagger}A\right]\}\right)+\\
+\frac{1}{2}\mathcal{L}^{s}\left(\Gamma=\left(\begin{array}{cc}
0 & 1\\
1 & 0
\end{array}\right),\{\left[B^{\dagger}B,A\right],A\}\right)+\\
+\frac{1}{4}\mathcal{H}^{s}\left(i\hbar\left[A^{\dagger}A,B^{\dagger}B\right]\right)\\
=\mathcal{L}_{1}^{s}(AB)-\mathcal{L}_{1}^{s}(BA)+\,\,\,\,\,\,\,\,\,\,\,\,\,\,\,\,\,\,\,\,\,\,\,\,\,\,\,\,\,\,\,\,\,\,\,\,\,\,\,\,\\
+\frac{1}{4}\mathcal{L}_{1}^{s}\left(B+\left[B,A^{\dagger}A\right]\right)-\frac{1}{4}\mathcal{L}_{1}^{s}\left(B-\left[B,A^{\dagger}A\right]\right)+\\
+\frac{1}{4}\mathcal{L}_{1}^{s}\left(A-\left[A,B^{\dagger}B\right]\right)-\frac{1}{4}\mathcal{L}_{1}^{s}\left(A+\left[A,B^{\dagger}B\right]\right)+\\
+\mathcal{H}^{s}\left(\frac{1}{4}i\hbar\left[A^{\dagger}A,B^{\dagger}B\right]\right).\label{eq:LL-comm}
\end{multline}

Note that the interaction of two noise terms leads to $6$ new noise
terms, but also to \emph{a coherent driving term}. This is expected
in light of \cite{Lie semigroups Dirr}, but its explicit form was not known.

\subsection*{Negative pre-fractors for Lindbladian terms}
Another issue requiring attention is the appearance of negative pre-factors to Lindbladian superoperators in both the Hamiltonian-Lindbladian and Lindbladian-Lindbladian commutators. Such negative rates normally correspond to re-coherence effects - rolling-back of decay processes \cite{neg-coef-1,neg-coef-2,neg-coef-3}, which are clearly non-Markovian. In fact, these negative pre-factors have been used to construct indicators of non-Markovianity \cite{Plenio review,neg-measure-1,neg-measure-2,neg-measure-3}. Note that the summation of these Lindbladian terms does not remove the difficulty.

The resolution of the dilemma is best viewed by considering the context in which these semigroup structure constants are often used, i.e. the BCH relations. As noted in the background review, the generator provided by the BCH relations matches the overall evolution only in the initial and final times. One therefore concludes, somewhat surprisingly, that superoperator describing the overall effect of a sequence of time-independent Markovian evolutions cannot itself be described by a time-independent Markovian evolution (which requires all-positive rates). This is in stark contrast to to unitary evolution, where the BCH amalgamation of time-independent unitaries, is, in itself, a time-independent unitary. Note that both time-dependent and time-independent Markovian evolutions are infinitesimally divisible quantum channels, as per \cite{Div. Q. Chann - Wolf and Cirac}.

\subsection*{Modification of Liouvillian evolution by coherent driving}

Consider a system (in the interaction frame) influenced by Markovian
noise. We shall drive it using a strong coherent impulsive drive $\Omega H$ for duration $t$, allow for free
evolution for duration $T\gg t$, during which the dominant effect is a Markovian noise $L$ and finally a counter-pulse, $-\Omega H$ of duration $t$ (ignoring Markovian effects during the pulses). Overall evolution is described by the superoperator
\begin{equation}\displaystyle
\exp\left(t\Omega\mathcal{H}\right)\exp\left(T\mathcal{L}_{1}\left(A\right)\right)\exp\left(-t\Omega\mathcal{H}\right).
\end{equation}
To first order in the generalized BCH series, this may be concatenated,
using eq. (\ref{eq:HL-comm}), as
\begin{multline}\displaystyle \exp\left(T\mathcal{L}_{1}\left(A\right)+2\frac{1}{2}tT\Omega\left[\mathcal{H},\mathcal{L}\right]\right)=\\
\exp\left(T\left(\mathcal{L}_{1}\left(A\right)-\frac{t\Omega}{2\hbar}\mathcal{L}_{1}\left(A+i\left[H,A\right]\right)+\frac{t\Omega}{2\hbar}\mathcal{L}_{1}\left(A-i\left[H,A\right]\right)\right)\right).
\end{multline}
Thus, it is clear that driving can shape the noise affecting the system (as opposed to reducing its rate). Note that this is a deeply different phenomena than Dynamical Decoupling \cite{Dynamical decoupling, Lidar review of DFS and DD}). In-fact, dynamical decoupling is wholly unrelated to the subject-matter of this report, as it is inherently a non-Markovian phenomena. Even in its most rudimentary bang-bang form, it requires a re-phasing period, decreasing the entropy of the open system in question. Therefore, dynamical decoupling is fundamentally unable to handle Markovian noise.

\subsection*{Pure coherent driving by interaction of Lindbladian terms}

Assume a system is subjected to one noise, $A$, followed by a period
with no noise, and later subjected to noise $B$. Furthermore, let
us assume both $A$ and $B$ are traceless. From eq. (\ref{eq:LL-comm})
it is clear that the combined effect, as detailed by the BCH series,
will include coherent terms. Surprisingly, in the case of single qubit
noise, there exist cases where, in the first order correction, all incoherent contributions disappear,
and $\left[\mathcal{L}^{s}\left(A\right),\mathcal{L}^{s}\left(B\right)\right]=\mathcal{H}^{s}\left(\frac{1}{4}i\hbar\left[A^{\dagger}A,B^{\dagger}B\right]\right)$.
Specifically in the following example:

\begin{equation}\displaystyle
A=\left(\begin{array}{cc}
1 & -4\\
3 & -1
\end{array}\right)\,\,\, B=\left(\begin{array}{cc}
-2 & 4\\
2 & 2
\end{array}\right),
\end{equation}

\begin{equation}\displaystyle
\left[\mathcal{L}^{s}\left(A\right),\mathcal{L}^{s}\left(B\right)\right]=\mathcal{H}^{s}\left(14\hbar\,\sigma_{y}\right).
\end{equation}

Note that incoherent terms remain in the zeroth order and second and higher orders (the latter may be removed at the limit by Trotterization).

\subsection*{Comments}

\subsubsection*{Comment re. Lamb shift}

Note that in eq. (\ref{eq:LL-comm}) we have terms of the form $A^{\dagger}A$
and $B^{\dagger}B$. These are well-known to be the Lamb shift terms.
However, their appearance here is surprising, as in the derivation
of the LKME, the Lamb shift terms are explicitly removed from the
Lindbladian and added to the Hamiltonian  (eq. (\ref{eq:HLS}), \cite{Textbook - Breuer,Rivas Huelga}
and elsewhere). Specifically,
they return in the noise-noise commutation term, as $\mathcal{H}^{s}\left(\frac{1}{4}i\hbar\left[A^{\dagger}A,B^{\dagger}B\right]\right)$,
which is of the same form as the commutation term of the Lamb-shift
factors added to the Hamiltonian, $\mathcal{H}^{s}\left(-\frac{i}{\hbar}\left[A^{\dagger}A,B^{\dagger}B\right]\right)$,
modifying its magnitude.

\subsection*{Superoperators as vectors in a higher Hilbert space}
Given a Markovian channel superoperator, one may directly reconstruct the Lindblad operators forming it by considering the superoperators associated with individual Lindblad and Hamitonian operators as vectors, collectively serving as a non-orthogonal basis for a higher ("super-super") dimensional object.

Going back to the general form of the LKME in eq. (\ref{eq:NonCanonicalLinbdblad}),
we shall choose the $L_k$ to be the traceless generators of $\textrm{SU}\crc{N}$, $\curly{S_k}_{k=1}^{N^2-1}$; and, since $H$ can be made traceless, we shall express it as a sum of the same generators, $H=\sum_k h_k S_k$. Let us denote by $\gamma_{j,k}$ a $\Gamma$ matrix (as in eq. (\ref{eq:NonCanonicalLinbdblad})) which is zero everywhere, except for the $\crc{j,k}$ element, which is $1$. One may show, by virtue of the linear independence of the generators, that the set of superoperators $B:=\curly{\mathcal{H}^{s}\crc{S_i},\mathcal{L}^{s}\crc{\gamma_{j,k},\curly{S_k}}}_{i,j,k=1}^{N^2-1}$ are linearly independent (but not orthogonal). Let us consider the super-operators as vectors in a higher (super-duper) Hilbert space. Now, we can view $B$ as a non-orthogonal basis (and a non-square matrix), to which we can construct the bi-orthogonal basis $G$, such that $G^{\dagger}B=\mathcal{I}$ (via the pseudo-inverse), with $G$ and $B$ spanning the same Hilbert space. Given the logarithm of a linear map $T$ (the issue of branches has been discussed in \cite{neg-measure-1}), one may project it onto the subspace spanned by $B$, via $G\log\crc{T}$. The resultant vector directly maps to the $h_k$ and $\gamma_{j,k}$ pre-factors of the operator's  Lindblad-form.

The above projection of the map's generator onto the space of Lindblad-form operators, which are infinitesimally divisible by construction, allows us to identify what remains in the orthogonal subspace as the non-divisible component.

\subsection*{Conclusion}

By deriving explicit expressions for the semigroup generator algebra,
we have provided a new tool to investigate the interaction of coherent
driving and Markovian noises, as well as the interaction of disparate
noise terms. Thus, we reveal a rich field of interactions, which
may be exploited to improve control of quantum systems.

Two unexpected results emerge. Firstly, that even pure coherent driving may be thus generated from interactions of
Markovian terms - a phenomena hitherto unknown. Second, we gain deeper understanding as to why the set of time-dependent Markovian maps is a proper super-set of time-independent Markovian maps (unlike unitary evolution, where this is not the case).
We believe that an extension of this approach to non-Markovian systems will enable new
insights into methods of controlling system decoherence, and will result
in a richer set of controls than have been known to-date.

This work was supported by an Alexander von Humboldt Professorship and the EU Integrating project SIQS.


\begin{thebibliography}{References}
\bibitem{alt veccing}One may alternatively transform $\rho$ into
a column by scanning by columns (as opposed to by rows), in which
case $\overrightarrow{L\rho R}=\left(\left(R^{T}\right)\otimes L\right)\vec{\rho}$
\end{thebibliography}
\end{document}